# iCellular: Define Your Own Cellular Network Access on Commodity Smartphones

Yuanjie Li[1], Haotian Deng[2], Chunyi Peng[2], Guan-Hua Tu[1], Jiayao Li[1], Zengwen Yuan[1], Songwu Lu[1]
[1] University of California, Los Angeles, [2] The Ohio State University

**Abstract**

Leveraging multi-carrier access offers a promising approach to boosting access quality in mobile networks. However, our experiments show that the potential benefits are hard to fulfill due to fundamental limitations in the network-controlled design. To overcome these limitations, we propose *iCellular*, which allows users to define and intelligently select their own cellular network access from multiple carriers. *iCellular* reuses the existing device-side mechanisms and the standard cellular network procedure, but leverages the end device's intelligence to be proactive and adaptive in multi-carrier selection. It performs adaptive monitoring to ensure responsive selection and minimal service disruption, and enhances carrier selection with online learning and runtime decision fault prevention. It is deployable on commodity phones without any infrastructure/hardware change. We implement *iCellular* on commodity Nexus 6 phones and leverage *Project-Fi*'s efforts to test multi-carrier access among two top US carriers: T-Mobile and Sprint. Our experiments confirm that *iCellular* helps users with up to 3.74x throughput improvement (7x suspension and 1.9x latency reduction *etc.*) over the state-of-art selection. Moreover, *iCellular* locates the best-quality carrier in most cases, with negligible overhead on CPU, memory and energy consumption.

## 1 Introduction

Mobile Internet access has become an essential part of our daily life with our smartphones. From the user's perspective, (s)he demands for high-quality, anytime, and anywhere network access. From the infrastructure's standpoint, carriers are migrating towards faster technologies (*e.g.*, from 3G to 4G LTE), while boosting network capacity through dense deployment and efficient spectrum utilization. Despite such continuous efforts, no single carrier can ensure complete coverage or highest access quality at any place and anytime.

In additional to infrastructure upgrade from carriers, a promising alternative is to leverage multiple carrier networks at the end device. In practice, most regions are covered by more than one carrier (say, Verizon, T-Mobile, Sprint, *etc.* in the US). With multi-carrier access, the device may intelligently select among carrier networks and improve its access quality. To this end, industrial efforts have recently emerged to provide 3G/4G multi-carrier access via universal SIM card, including Google Project Fi [29], Apple SIM [16], and Samsung e-SIM [27]. The ongoing 5G standardization also seeks to integrate heterogenous network technologies [39].

However, our study shows that the full benefits of multiple carrier access can be limited by today's cellular design. We examine Google *Project-Fi* with two carriers (T-Mobile and Sprint), and discover three problems, all of which are independent of implementations (§3): (P1) the anticipated switch never occurs even when the serving carrier's coverage is really weak; (P2) the switch takes rather long time (tens of seconds or minutes) without service availability; and (P3) the device fails to choose the high-quality network (*e.g.*, selecting 3G with weaker coverage rather than 4G with stronger coverage).

It turns out that, these problems are rooted in the conflicts between legacy 3G/4G roaming design and user's multi-carrier access requests. With the single-carrier scenario in mind, the 3G/4G design places the controllability of carrier access to the network side. Roaming to other carriers is not preferred unless the home carrier is unavailable. As a result, today's carrier selection mechanism (i.e., PLMN selection) passively monitors other carriers after losing home carrier service, and selects the carrier based on pre-defined roaming preference given by the serving carrier network [13, 14]. Although viable in the single carrier case, this design limits user's ability to explore multiple carriers. The user could miss the high-quality carrier network, delay the switch with redundant carrier scanning, and get stuck in the low-quality carrier.

While this problem may be solved in future architecture design (*e.g.* 5G), it takes years to accomplish. Instead, we seek to devise a solution that works in today's 3G/4G network, in line with ongoing industrial efforts,

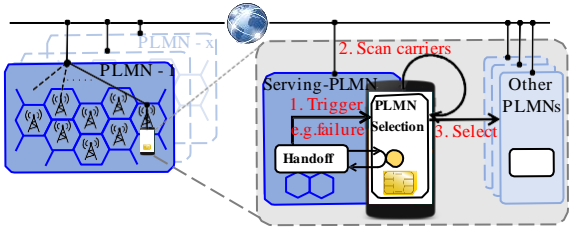

Figure 1: **Multi-carrier network access (left) and inter-carrier switch via PLMN selection (right).**

*e.g.* Google *Project-Fi*, Apple SIM and Samsung e-SIM. Specifically, we address the following problem: *can we overcome the design limitations of legacy 3G/4G roaming, without modifying phone hardware and 3G/4G network infrastructure?* Our study yields a positive answer.

We propose *iCellular*, a phone-side service to let users define their own cellular network access. Different from the traditional network-controlled roaming, *iCellular* enhances the user's role in multi-carrier access. It offers users high-level APIs to customize the access strategy. *iCellular* is built on top of current 3G/4G mechanisms at the device, but applies cross-layer adaptations to ensure responsive multi-carrier access with minimal disruption. To help users make proper decisions, *iCellular* exploits online learning to predict heterogenous carrier's performance. It further safeguards access decisions with fault prevention techniques. We implement *iCellular* on commodity phone models (Nexus 6) and assess its performance with *Project-Fi*. Our evaluation shows that, *iCellular* can help users gain 3.74x throughput improvement and 1.9x latency reduction on average by selecting the best carrier network. Meanwhile, *iCellular* has negligible impact on user's data service and OS resource utilization (less than 2% CPU usage), approximates the lower bound of responsiveness and switch disruption, and shields user strategies from decision faults.

The rest of the paper is organized as follows. §2 introduces background. §3 presents experimental findings on issues of multi-carrier access, and uncovers the root causes. §4, §5, and §6 describe the design, implementation and evaluation of *iCellular*. §7 discusses other alternatives, and §8 compares with the related work. §9 concludes the work.

## 2 Mobile Network Access Premier

A cellular carrier deploys and operates its mobile network to offer service to its subscribers. Such network is defined as a public land mobile network (PLMN). Each PLMN is divided into multiple geographical regions called cells. In reality, one location is likely covered by multiple cells within one PLMN and multiple PLMNs (*e.g.* Verizon, AT&T, T-Mobile, Sprint).

**Single-carrier network access.** Today's cellular network is designed under the premise of single-carrier network access. A user device is supposed to gain access directly from the home PLMN. It obtains radio access from the serving cell and further connects to the core carrier network and the external Internet, as shown in the left plot in Figure 1. When the current cell can no longer serve the user device (*e.g.*, out of its coverage), the device will be migrated to another available cell within the same PLMN. This is called as handoff. For each single carrier, there exists two types of handoffs: handoff within the same radio access technologies (RAT, *e.g.*, 4G→4G) and inter-RAT handoff (*e.g.*, 3G→4G).

**Roaming between carriers.** When the home PLMN cannot serve its subscribers any more (*e.g.*, in a foreign country), the device may roam to other carriers (visiting PLMN networks). In cellular network, this is enabled by PLMN Selection between carriers [14]. It supports both automatic mode (based on a pre-defined PLMN priority list) and manual mode. As shown in the right plot of Figure 1, once triggered by certain events (*e.g.* no home PLMN service or handoff failure), PLMN selection should first scan the available carriers, and then select one based on pre-defined criteria (*e.g.* preference) or user manual operation. If the device determines to switch, it will deregister from the current carrier network, and register to a new one. In this process, the network service may be temporarily unavailable. It is acceptable because inter-carrier switch is supposed to happen rarely, thus having limited impact on user's data/voice usage.

**Multi-carrier access with universal SIM card.** Recent industrial efforts aim at providing user access to multiple cellular carriers with single SIM card. This includes Google Project Fi [29], Apple SIM [16], and Samsung e-SIM [27]. With installation of the SIM card, the user device has access to multiple cellular carriers (*e.g.* T-Mobile and Sprint in *Project-Fi*). Since only single cellular interface is available, each time the device can only use one of the carriers. Similar to roaming, the switch between carriers is also based on PLMN selection.

## 3 Multi-carrier Access: Promises & Issues

In this section, we run real-world experiments to justify the benefits of accessing multiple carriers, and point out the limitations of the today's mechanism to motivate our design. The limitations we identify are independent of any implementations. They are rooted in 3G/4G design.

**Methodology:** We conduct both controlled experiments and a one-month background user study using two Nexus 6 phones with *Google Project-Fi* [29]. *Project Fi* is the only workable system in the market to offer users with access to two top-tier US carrier partners: T-Mobile and Sprint. It was released to the US in May 2015 and is still at its early stage (invitation only). It leverages the standard PLMN selection procedure and works with the commodity phone (so far, only Nexus 6 is supported).

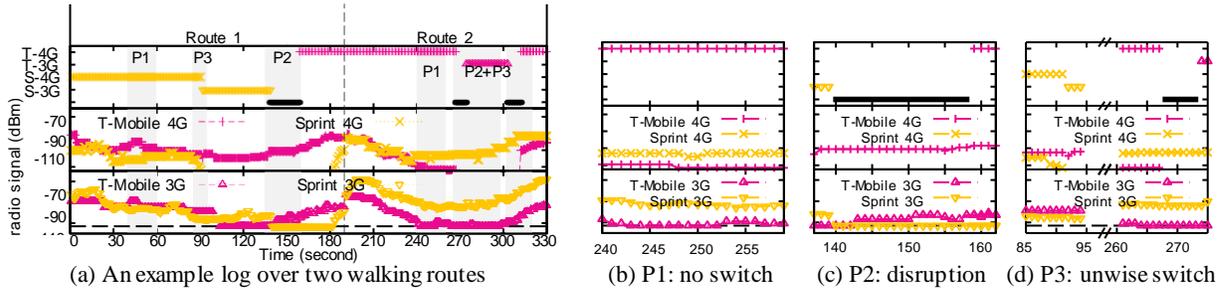

Figure 2: An example log for serving carriers and networks and three problematic instances through *Project-Fi*.

In each controlled experiment, we use a Nexus 6 phone with a *Fi*-SIM card and walk along two routes in the campus building at UCLA or OSU. We walk slowly (< 1 m/s) to record the serving carrier ("T" for T-Mobile, "S" for Sprint) and its network type (4G or 3G) per second. In the meanwhile, we carry other accompanying phones to record the radio signal strength of each access option (T-4G, T-3G, S-4G, S-3G). We run 10 times and similar results are consistently observed in all the tests. In the user study (07/31/15 to 09/02/15), we use the *Fi*-enabled phone as usual and collect background user and cellular events with MobileInsight, an in-phone cellular monitoring tool [4]. Here, we present two controlled experiments as motivating examples, and describe more experiments and the user study results in §4 and §6. The user study verifies that the problems in the examples are common in practice.

### 3.1 Motivating Examples

**Merits of multi-carrier access:** we first verify that exploring multiple carriers is indeed beneficial for service availability and quality. Figure 2a shows the controlled experiment results over two routes. On the first route [0s,190s], Sprint gradually becomes weaker and then completely fades away but its dead-zone is covered by T-Mobile; On the second route [190s, 330s], in contrast, Sprint offers stronger coverage, even at locations with extremely weak coverage from T-Mobile. Clearly, multi-carrier access indeed helps to enhance network service availability and quality, at least boosting radio coverage. We later demonstrate that such enhancement can greatly improve service performance (*e.g.*, data speed) and user experience. For example, in [160s, 180s], the phone switches to T-Mobile and sustains radio access while Sprint is not available. Recent efforts (*e.g.*, *Google Project-Fi*, *Apple SIM*) make a promising step.

However, the benefits and potentials of better carrier access options have not been fully achieved. These examples also reveal three problems, which are common in practice and rooted in today's 3G/4G design.

**P1. No anticipated inter-carrier switch.** It is desirable for the device to migrate to better network access from other carriers available once it experiences lousy channel from the currently serving carrier. However, the experiments show that the mobile device often gets stuck in one carrier's network and miss the better network access, for example, during [40s,60s] and [240s, 260s] in Figure 2. As shown in Figure 2b, T-Mobile suffers from extremely weak radio coverage (< −130 dBm in 4G and < −110 dBm in 3G), but the phone never makes any attempt to move to Sprint, regardless of how strong Sprint's radio signal is. As a result, the device misses the opportunity to improve network access quality. Moreover, we find that the expected switch only occurs until its access to the original carrier (here, T-Mobile) is completely broken. This is rooted in the PLMN selection practice which triggers inter-carrier switch only when the serving carrier fails. In other words, the device has to become out of service in this case, which is against the intention to promote access with more carrier choices.

**P2. Long switch time and service disruption.** Even when inter-carrier switch is eventually triggered, it may interrupt access for tens of seconds or even several minutes (see Figure 7 for the user study results). In this example (Figure 2c), the mobile phone starts Sprint→T-Mobile roaming at the 140th second but it takes 17.3 seconds to finally gain access to T-Mobile 4G. This duration is exceptionally longer than the typical handoff time (likely, several seconds) [46]. Such interruption likely tears down ongoing data services. We look into the event log (Figure 3) to figure out why the switch is so slow. It turns out that most of switch time is wasted on an *exhaustive* scan of all the possible cells, including AT&T and Verizon cells around. In this example, it spends 14.7s on radio band scanning and 2.6s on completing the registration (attachment) to the new carrier (here, T-Mobile). One thing worth noting is that such heavy overhead is not an implementation glitch (done by Google); Instead, it is rooted in the standard design which selects a new PLMN only after an exhaustive scan [14]. However, we will disclose that such long delay is unnecessary and can be reduced without scarifying inter-carrier selection.

**P3. Unwise decision and unnecessary performance degradation.** We observe that the device fails to migrate to the better choice, thus unable to achieve the potential benefits of multi-carrier access. The phone often moves to 3G offered by the same carrier, rather than the 4G network from the other carrier which likely yields higher bandwidth and faster speed. Figure 2d illustrates two instances where the other carrier even provides much stronger 4G access. After entering into an area without Sprint 4G at the 91st second, the device moves to Sprint 3G network, despite stronger radio quality from T-Mobile 4G. It indicates that the intra-carrier handoff is preferred to the inter-carrier switch. However, such preference prevents the inter-carrier switch from taking effect in reality. Even worse, such obstacle still exists when its network access to the original carrier has been shortly disrupted. For instance, in [267s, 273s], the original carrier (T-Mobile 3G) is still chosen. In this case, T-Mobile 4G and 3G networks almost have no coverage. In other words, the mobile device acts as single-carrier phone in most cases, even with the multi-carrier access capability. Inter-carrier switch rarely happens as expected.

| Time | Event | |
|---|---|---|
| 11:19:57.414 | Out-of-service. Start PLMN search | |
| 11:19:57.628 | Scanning AT&T 4G cell 1, unavailable | RF band scanning: 14.7s |
| 11:19:57.748 | Scanning AT&T 4G cell 2, unavailable | |
| ... | ... | |
| 11:20:11.788 | Scanning Verizon 4G cell 1, unavailable | |
| 11:20:12.188 | **Scanning T-Mobile 4G cell 1, available** | Network registration: 2.6s |
| 11:20:12.771 | Attach request (to T-Mobile 4G) | |
| 11:20:14.788 | Attach accept | |

**Figure 3: Event logs during P2 (disruption) of Fig. 2c.**

## 3.2 Root Causes In Net-controlled Access

We further explore the root causes of the problems by analyzing 3G/4G design. It turns out that, all the problems are a result of *network-controlled* PLMN selection in 3G/4G, which was a viable choice for single-carrier usage, but not appropriate with multi-carrier access.

Today's cellular networks are designed under the premise of single-carrier access. PLMN selection is no exception. While roaming to other carrier networks (visiting PLMNs) is allowed, it is not preferred by the home PLMN unless it fails to offer network access to its subscribers. With this in mind, the 3G/4G design mandates the phones to choose carriers with the following PLMN selection procedure [13, 14]: (1) *passive triggering/monitoring:* when being served by one PLMN network, the device should not monitor other carriers or trigger the selection until the current network fails (*i.e.*, out of coverage); (2) *network-controlled selection:* the device should choose the visiting network based on the PLMN preferences predefined by the home carrier, which are stored in the SIM card. (3) *hard switch:* the device should deregister from the old network, and register to the new one.

Such design places the controllability of multi-carrier access on the network side, though it is executed on the phone side. It determines when to switch, what to select and how to execute, based on the pre-defined criteria or configurations from the network side, regardless of useful information available on the phone side. For example, the phone is able to probe other available carriers and determine whether inter-carrier switch is beneficial; It can trigger a preferable switch before it loses access; The phone has historical information and user preference which can help to filter out unnecessary scanning and select the preferable access faster; The phone is able to minimize disruption if it leverages the app context information (whether the apps are running). End intelligence is a necessity to make high-quality multi-carrier access while the legacy design uses network intelligence only. Moreover, network intelligence is insufficient for multi-carrier access. Each individual carrier by no means has a global view, as it lacks information of other carriers which are available to end devices. When the phone has a global view, the handoff within a single carrier may not be preferred to those inter-carrier switch. Note that, the legacy design is rooted in the telcomm principle of "*smart core dumb end*"; However, today's end devices are not dumb any longer, with proliferation of smartphones.

**Problem statement:** we investigate an alternative multi-carrier access solution that address above problems. Different from the network-controlled approach, the solution should give users more power to define their own cellular carrier access. It should allow responsive decisions, without missing the desired carrier network or making unwise decisions (P1 and P3). It should not disrupt available network service, or incur long unavailability in switch (P2). We seek a solution that can be readily deployable on commodity smartphones, and work coherently with existing 3G/4G network infrastructure.

## 4 *iCellular* Design

We now present *iCellular*, a device-side software-based solution that facilitate users to define their own cellular network access. Figure 4 gives an overview. In brief, *iCellular* enhances the users' role in every step of inter-carrier switch, including the triggering/monitoring, decision and switch execution. It provides high-level APIs for users to customize selection strategy (§4.1). To be incrementally deployable, we build *iCellular* on top of existing mechanisms from phone's cellular interface. To guarantee the responsiveness and minimal-disruption, *iCellular* applies adaptations over existing mechanisms (§4.2 and §4.3). To help users make wise decisions, *iCellular* offers a online learning service to predict network performance (§4.4), and protects users from decision faults (§4.5). To achieve adaptation, prediction and decision fault prevention, *iCellular* incorporates with real-time cellular event feedbacks. Different from approaches using additional diagnosis machine (*e.g.* QXDM [42]) or software-defined radio (*e.g.* LTEye [35]), we develop an in-phone mechanism to gather realtime cellular events (§4.6, cellular events are summarized in Table 1).

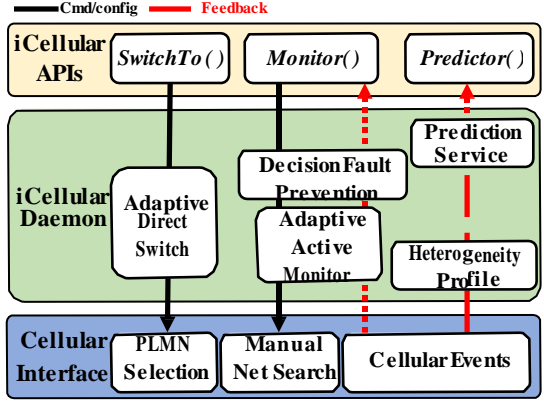

Figure 4: *iCellular* system architecture.

| Function | Method | Cellular Events | Type |
|---|---|---|---|
| Active monitor (§4.2) | Disruption avoidance | Paging | Meas |
| | | Paging cycle | Config |
| | Minimal search | Radio meas | Meas |
| | | RRC SIB 1 | Config |
| Prediction service (§4.4) | QoS profile | EPS/PDP setup | Config |
| | Radio profile | RRC reconfig | Config |
| Decision fault prevention (§4.5) | Access control | RRC SIB1 | Config |
| | Interplay with net mobility | Cell reselection in RRC SIB 3-8 | Config |
| | Function completeness | GMM/EMM location update | Config |

Table 1: Cellular events used in *iCellular*.

### 4.1 *iCellular* APIs

*iCellular* allows users to control their cellular access strategy through three high-level APIs: Monitor(), Predictor() and SwitchTo(). We use a simple example to illustrate how they work. Consider a user who has access to T-Mobile and Sprint 3G/4G networks, and would like to choose in the network with minimal radio link latency. To do so, the user first initiates an *active monitor* by calling Monitor() function, and specifying the list of the carrier networks s/he is interested in:

monitor = Monitor(["T-4G","T-3G","S-4G"]);

Different from the PLMN selection, active monitoring allows users to scan carriers even if their current carrier network is serving them. This would prevent users from missing the better carrier network (P1 and P3 in §3).

To choose the target carrier network, the user may want to learn each network's performance (latency in this example). But without registration to other carriers, it cannot directly probe other carrier network's performance at runtime. To facilitate users make decisions, *iCellular* defines a *prediction service*, which allows user to specify the metric it is interested, and learns a prediction model based on heterogeneity profiles and runtime monitoring results. The following code shows how user can initiate a latency predictor[1]:

predictor = Predictor("Latency");

The last step is to define the access selection strategy and perform the switch if applicable. To let users make responsive decisions, *iCellular* let user strategy be triggered by the latest and even partial search results. To do so, the user should overload an *event-driven* decision callback function. Users are given the runtime monitoring results of available carrier networks. Optionally, the user can use the predictor to help determine the target carrier network. The user can call SwitchTo() function to perform the inter-carrier switch. Different from the PLMN selection, the SwitchTo() performs *direct* inter-carrier switch, and minimizes the service disruption time (P2 in §3). The following code shows a strategy that minimizes latency and ensures satisfying radio quality (greater than -100dBm):

```
def decision_callback(monitor):
  min_latency = inf; target = null;
  for network in monitor:
    latency = predictor.predict(network);
    if network.rss > -100dBm
    and latency < min_latency:
      min_latency = latency; target = network;
  SwitchTo(target);
```

We next elaborate how *iCellular* realizes these abstractions in an incrementally deployable way, while still guaranteeing responsiveness and minimal-disruption.

**Access to multiple carriers simultaneously?** The readers may notice that *iCellular* does not provide API to simultaneous registration to multiple carriers. This may help utilize all available cellular networks, and has been extensively discussed in MPTCP [43, 47] and WiFi [22, 33]. Unfortunately, it cannot be supported by either 3G/4G network infrastructure or device. On the network side, the registration on one carrier's 3G/4G network would result in de-registration from other carriers by design. Today's roaming architecture maintains user identify and location state information in a *single* home subscriber server (HSS). Whenever the device registers on a new carrier network, the 3G/4G design requires the HSS to delete old location area context, and sends a *location cancellation* command to the old location area controller to deregister the user [9, 19]. On device side, most phones only have single cellular interface, and do not support storing multiple network states. While there exist phones with dual SIM cards [2], they do not scale to more carriers. One could also propose clean slate designs, which however would require long-term deployment. While we appreciate the benefits of simultaneous cellular access, we prefer our solution to be readily deployable.

---
[1]In the real implementation, the user should provide a callback to let *iCellular* monitor this metric.

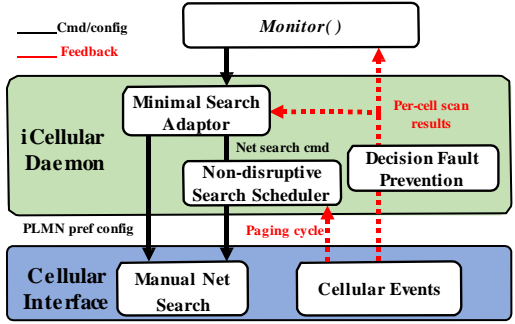

Figure 5: Adaptive active monitoring in *iCellular*.

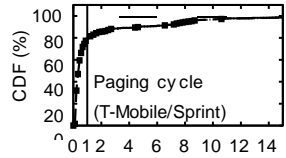

Figure 6: Cell scan time.

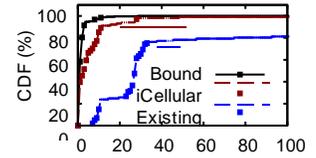

Figure 7: Switch time.

### 4.2 Adaptative Monitoring

To achieve the active monitoring function Monitor(), *iCellular* should initiate the carrier network search even when the current network is still available. For this purpose, the only available mechanism for user in phone is the manual network search [14]. It was designed to let users manually scan all available carriers. Once initiated, the device scans neighboring carrier's frequency bands, extracts the network status from the broadcasted system information block (SIB), and measures their radio quality. To be incrementally deployable, we choose to realize Monitor() on top of manual network search.

However, naive manual search does not satisfy min-disruption or responsiveness properties. First, scanning neighbor carriers may disrupt the network service. The device has to re-synchronize to other carriers' frequency bands, during which it cannot exchange traffic with current carrier network. Second, it is *exhaustive* to all carriers by design. Even if the users are not interested in some carriers (*e.g.* no roaming contract), this function would still scan them, which can delay the user decision and wastes more power. To address them, *iCellular* introduces cross-layer adaptors for both problems (Figure 5):

**Disruption avoidance:** to avoid disruption on user data service, *iCellular* schedules scanning events only when the device has no traffic delivery. This requires *iCellular* to monitor the uplink and downlink traffic activity. Although the uplink traffic activity can be directly known from device itself, the downlink traffic status is challenging to predict. They may arrive when the device has re-synchronized to other carriers' cells. If so, these traffic reception could be delayed or even lost.

*iCellular* prevents this with the low-level cellular event feedback. We observe that in the 3G/4G network, the downlink data reception is regulated by the periodical paging cycle (*e.g.* discontinuous reception in 4G [11, 44]). To save power, the 3G/4G base station assigns inactivity timer for the device. The device periodically wakes up from sleep mode, monitors the paging channel to check downlink data availability, and then moves to sleep mode if no data is available. *iCellular* gains the this cycle configuration from the radio resource control (RRC) message, and schedules scanning events only when device is in the sleep mode. Figure 6 shows our one-month observation of 4G per-cell search time in one mobile device with *Project-Fi*. It shows that, 79.2% of cells can be scanned in less than one paging cycle. For other cells, more cycles are needed to finish the scanning.

**Minimal search:** instead of exhausting all carrier networks, *iCellular* adapts manual search to only scan those specified by user. Given the user-specified carrier networks in Monitor(), *iCellular* first configures the cellular interface to let the manual network search scans these carriers first. This is achieved by assigning these carriers with highest PLMN preferences. Then *iCellular* initiates the manual search, and listens to the cellular events to see which carrier is being scanned. These events include the per-cell radio quality measurements, and its system information block with PLMN. When *iCellular* detects that the device has finished scanning of the user-specified carriers, it terminates the manual network search function.

**Monitoring-decision parallelism:** sometimes the user can determine the target carrier network without complete monitoring results. For example, if the user prefers 4G, it can decide to switch whenever 4G is reported, without waiting for 3G results. To support this, *iCellular* allows users to make decisions with partial results, thus further accelerating the decision. This is realized with *iCellular*'s event-driven API design: instead of waiting all scanning results, *iCellular* triggers the decision callback whenever new results are available.

### 4.3 Direct Inter-carrier switch

*iCellular* aims at minimizing the disruption time when switching between carriers. To be incrementally deployable, *iCellular* still reuses the PLMN selection, but applies adaptation to minimize the switch disruption. This is doable because *most service disruption time are caused by necessary frequency band scanning* (§3). With the active monitoring function, *iCellular* does not need to scan the carrier networks in switch. More specifically, given a target carrier network from SwitchTo(), *iCellular* achieves the direct switch by configuring the target carrier with highest PLMN preference. Then it triggers manual PLMN selection to the target carrier network. This way, the device would directly switch to the target without unnecessary scanning.

We next compare *iCellular*'s switch time with lower bound in 3G/4G. In 3G/4G network, switching to another network requires at least de-registration from the old network (detach), and registration to the new network (attach). According to [12], the detach time is negligible, since the device can detach directly without interaction to the old carrier network. So the minimal disruption time in switch is equal to the attach time, *i.e.* $T_{switch,min} = T_{attach}$. For *iCellular*, no extra attempts to other carrier networks are performed. But since it is on top of PLMN selection, the scanning of target carrier still exists. So the inter-carrier switch time is

$$T_{switch,iCellular} = n_t T_t + T_{attach} = T_{switch,min} + n_t T_t \quad (1)$$

where $n_t$ and $T_t$ are the cell count and per-cell scanning time for the target carrier network, respectively. Compared with the attach time, this extra overhead is usually negligible in practice. Figure 7 verifies this with our one-month background monitoring results in *Project-Fi*. It shows that, *iCellular* indeed approximates the lower bound, despite this minor overhead.

### 4.4 Prediction for Heterogeneous Carriers

To decide which carrier network to switch to, the users may gather performance information for each carrier network. Ideally, all the information should be collected at runtime to facilitate accurate decisions. Unfortunately, this is not available with registration to single carrier network. Unregistered carrier networks would not allocate any resource to the users, and the device cannot exchange information with these networks. Only the radio quality/load measurements and basic system information is available for unregistered carrier networks.

Even so, one may wonder if it is sufficient to determine the target carrier network with radio quality/load measurements only. In fact, this is the *de facto* approach for 3G/4G handoffs within a single carrier network [11, 13]. However, besides radio quality/load, the heterogeneity between carrier networks also plays a critical role for differentiating their performances. Different cellular carriers may deploy heterogeneous radio technologies (*e.g.* 3G UMTS VS. EvDo) and operation rules even for the same radio technology (*e.g.* QoS and radio configurations). Compared with the handoff within a single carrier, in inter-carrier switch such heterogeneity may have a larger impact on performance. In §6.1, we will show that radio-only strategies are not sufficient to help users select satisfying carrier networks.

Given this fact, *iCellular* chooses to help users predict carrier's performance based on both radio measurements and heterogeneity information. This approach is feasible

|  | Profile | Sprint | | T-Mobile | |
|---|---|---|---|---|---|
|  |  | Value | Probability | Value | Probability |
| QoS | Traffic class | Background | 100% | Interactive | 97.5% |
|  | Delay class | 4 (best effort) | 100% | 1 | 100% |
|  | Max DL rate | 200Mbps | 100% | 256Mbps | 100% |
|  | Max UL rate | 200Mbps | 100% | 44Mbps | 100% |
|  | DL GBR | N/A (best effort) | 100% | N/A (best effort) | 100% |
|  | UL GBR | N/A (best effort) | 100% | N/A (best effort) | 100% |
| Radio | TDD UL/DL | Type 1 | 100% | N/A | |
|  | Paging cycle | 1s/2s[a] | 81.5% | 1s | 99.4% |
|  | Handoff priority | 2/3/6[a] | 100% | 2/3/6[a] | 100% |
|  | Handoff threshold | -120dBm | 100% | -120dBm | 100% |

**Table 2: Predictability of heterogeneity profiles.**

[a]Varies between 4G frequency bands, but invariant within the band.

because *the radio technology and operation heterogeneity between carriers are predictable*. We validate this assumption by analyzing a 1-month background user study log, which includes 13257 radio configuration and 939 QoS configuration messages, and covers 136 4G and 76 3G base stations. Table 2 lists the predictability of some parameters from this log. For each parameter, we choose the one with highest probability, and shows it occurrence probability. It shows that, most QoS and radio configurations are invariant of time and location. Some parameters (*e.g.* priority) may vary between base stations, they are predictable on per-cell granularity. By profiling these heterogeneity information, *iCellular* can use them for later predictions. To this end, *iCellular* learns the prediction model as follows:

**Heterogeneity profiling:** when registered in different carrier networks, *iCellular* collects the cellular events, extract the configurations and aggregates them by time and location. Currently *iCellular* focuses on two types: (1) QoS profile from the EPS(4G)/PDP(3G) context, which includes the delay class and peak/maximum throughput; (2) radio parameters from the RRC configuration message, which includes the physical and MAC layer configurations. To reduce the training sample dimension, we exclude fields unrelated to the performance based on the domain-specific knowledge (*e.g.* temporarily identifiers, timestamps and security functions).

**Online predictor training:** *iCellular* uses regression tree algorithm [38] to learn a predictor for user-specific metric from Predictor(). The predictor is represented as a tree, with each interior node as a test condition over radio measurements or profile fields. When user calls Predictor.predict(), *iCellular* takes measurements and heterogeneity profile as input, traverse the tree to the leaf and returns the estimation.

To construct the regression tree, *iCellular* starts from the root, and finds the field (measurement or profile) that best splits the samples by minimizing the impurities in the two child nodes. This field would be used as the test criteria for the current node. Then we move on to its children, and recursively add nodes until further adding nodes do not give extra information. Then we assign each leaf node an estimated user-specified metric value based on regression. We use an online version [21], so that it can be trained with incremental sample collection from *iCellular*, and improve the prediction accuracy by accumulating more samples. Moreover, since the heterogeneity profile is highly predictable, *iCellular* further optimizes the algorithm by caching the branches whose parent node tests heterogeneity profiles. In both training and query phase, this reduces computations.

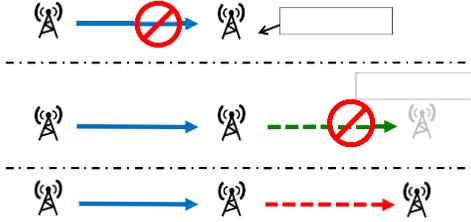
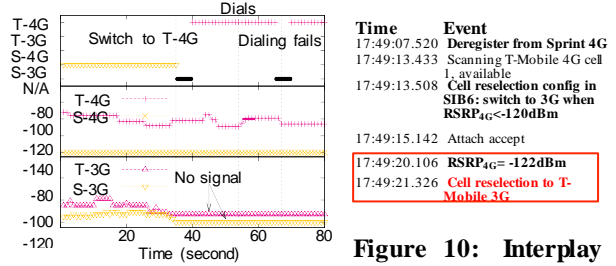

Figure 8: Three types of improper switch decisions.

Figure 9: Switch to a network with no voice support.

Figure 10: Interplay between user and network's mobility.

### 4.5 Decision Fault Prevention

Letting users customize access strategy can be a double-edged sword. Users may make faults and cause unexpected service disruption and/or undesired switch. Figure 8 shows three categories of user decision faults:

**Forbidden access:** some carrier network may be temporarily inaccessible. For example, our user study reports that a Sprint 4G base station experiences 10-min maintenance, during which the access baring option the the RRC system block is enabled.

**Switch to carriers with incomplete service:** in some scenarios, the target carrier network cannot provide complete data/voice services. Figure 9 shows one instance from our user study. T-Mobile provides voice service with circuit-switch-fall-back, which moves user to 3G and utilizes circuit-switch service for the call. But there exist areas not covered by T-Mobile 3G. In this scenario, the user in Sprint 4G should not switch to T-Mobile 4G, which cannot offer voice service without 3G network.

**Incoordination with carrier's mobility rules:** the user selection may not be honored by individual carrier's handoff rules. Figure 10 reports one instance from our user study: the user under Sprint 4G may determine to switch to one T-Mobile 4G. But under the same condition, T-Mobile's mobility rules (*e.g.* cell re-selection [13]) determine to switch its 4G users to its 3G. In this case, user's decision to T-Mobile 4G is improper, because the final target network (T-Mobile 3G) is not preferred, and such switch incurs unnecessary disruption.

To prevent users' decision faults, *iCellular* chooses to safeguard user decisions strategies with fault-prevention mechanisms (Figure 5). Given the runtime measurements and heterogeneity profiles, *iCellular* estimates whether each carrier network has one of the problems above. If so, this carrier network would be excluded from the monitoring results to user decision callback. This prevents user from switching these carrier networks. To detect forbidden access, *iCellular* checks the access control list from RRC SIB 1 [11]. To detect function incompleteness, *iCellular* checks the profiled data/voice preference configuration from registration/location update messages [12]. To prevent incoordination with network-controlled handoff, *iCellular* profiles each carrier network's mobility rules from RRC configuration message [11,13], and predicts if further handoff would be initiated by network with measurements.

### 4.6 Cellular Events Collection

As shown in §4.2-§4.5, *iCellular* relies on low-level cellular events to perform cross-layer adaptations over existing mechanisms, predict the carrier network performance and potential switch faults. The cellular events include the signaling messages between device and network, and radio quality/load measurements. Table 1 summarizes the events required by *iCellular* functions. Note that some events (*e.g.* paging) should be extracted at realtime for feedbacks. Unfortunately, getting cellular events on commodity phone at realtime is not readily available today. These events are not exposed to mobile OS or apps. There exist commercial tools (*e.g.* QXDM [42]) and research projects (*e.g.* LTEye [35] and CellIQ [30]) to extract them. But both require external platform (*e.g.* laptop) to connect to the mobile device, which limits the device's flexible movement, and cannot satisfy *iCellular* realtime demand. To this end, we develop an in-phone solution by adapting the existing cellular diagnostic mode. We enable the diagnostic mode on the phone, modify the the virtual device for it, and redirect the cellular events from the USB port to the memory. This way, no external platform is needed.

## 5 Implementation

We implement *iCellular* on Motorola Nexus 6, a commodity smartphone released in Oct 2014. It runs Android OS 5.1 over Quad-core 2.7 GHz Krait 450 (CPU) and

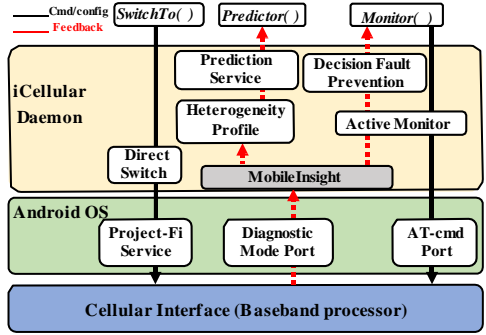

**Figure 11: Overview of *iCellular* implementation.**

Qualcomm Snapdragon 805 (cellular interface). It supports 4G LTE, 3G HSPA/UMTS/CDMA and 2G GSM. To activate access to multiple cellular networks, we installed *Project-Fi* SIM card on Nexus 6, which supports T-Mobile and Sprint 3G/4G.

Figure 11 illustrates the system implementation. *iCellular* runs as a daemon service on a rooted phone. To enable interaction with the cellular interface, we first activate the baseband processing tools (in bootloader), and turn on the diagnostic mode [3] and AT-command interfaces. To facilitate implementation, we further develop MobileInsight [4], a Python package that supports analysis of major 3G/4G protocols, including the radio resource control (RRC), mobility management (GMM/EMM) and session management (SM/ESM).

***iCellular* APIs (§4.1):** *iCellular* implements them under the Analyzer, an event-driven interface in MobileInsight for the cellular message analysis. It accepts <filter, callback> pair at runtime, extracts corresponding cellular events, and triggers the callbacks for them. This makes event-driven decision strategy callback implementation straightforward. To define a selection rule, the users can develop a script with APIs in Python. Then *iCellular* will load them at runtime.

**Adaptative active monitoring (§4.2):** we implement the Monitor() with manual search and adaptations. Our implementation initiates the search with an ATquery command AT+COPS=?. The non-disruption and minimal search adaptations are implemented as an Analyzer for cellular events in Table 1.

**Adaptive direct switch (§4.3):** we implement SwitchTo() method on top of PLMN selection, with dynamic adaptations for direct switch. Ideally, this can be achieved with the AT set command AT+COPS=manual,carrier,network. However, this command is forbidden by Nexus 6's cellular interface. So we enable an alternative approach: we change the preferred network type through Android's API setPreferredNetworkType, and change carrier with *Project-Fi*'s secret code (34777 for Sprint, 34866 for T-Mobile). Admittedly, this approach may incur extra switch overhead but still tolerable (§6.2).

**Prediction for heterogenous carriers (§4.4):** we implement Predictor() in two steps. First, we implement the heterogeneity profiling under Analyzer, with filters for the radio measurements, RRC configuration and QoS profiles. Then we implement the incremental regression tree algorithm as another Analyzer for the profiling and user-specific metric monitoring function.

**Decision fault prevention (§4.5):** the fault prevention function is implemented as a shim layer between active monitoring and user APIs. It predicts the potential switch failures based on monitoring results and heterogeneity profiling, and exclude the unreachable carrier networks from the monitoring results. We further add a runtime checker in SwitchTo() function, and prevent users from selecting carriers not in the scanning results.

**Cellular events collection (§4.6):** we use the built-in realtime cellular monitors from MobileInsight. We modify the diagnostic mode port (/dev/diag), and redirect the events to the phone memory.

## 6 Evaluation

We evaluate *iCellular* in two dimensions. We first present the overall performance improvement achieved by *iCellular* with smart multi-carrier access, and then show *iCellular* satisfies various design properties in §4. All the evaluations are conducted on the commodity Nexus 6 phones implementing *iCellular*, and tested in two cities of Los Angeles (west coast) and Columbus (Midwest), mainly around two campuses.

### 6.1 Overall Performance

We choose four representative apps to evaluate *iCellular*'s performance: downloading SpeedTest for bulk file transfer, interaction latency for small volume traffic (web), video (Youtube) and VoIP (Skype). *iCellular* can support customizing access strategy using other metrics (*e.g.* minimizing billing, see Appendix A). We evaluate each app with key quality-of-experience metrics whenever possible, *i.e.*, downlink speed for SpeedTest, page loading time for Web [15], video suspension time for Youtube [36], latency for Skype [31]. We test with the following traffic: web via loading Yahoo News webpage, Youtube via watching a 10-min HD (auto rate) video, Skype via making a 2-min voice call. Before each run, we clear the caching for Web and Youtube (no need for Speedtest and Skype). The details to collect app performance metrics are given in Appendix B. We run both walking-mobility and static tests. Along the walking routes, we uniformly sample locations and perform additional static tests. We run at least 5 times and use the median value for evaluation.

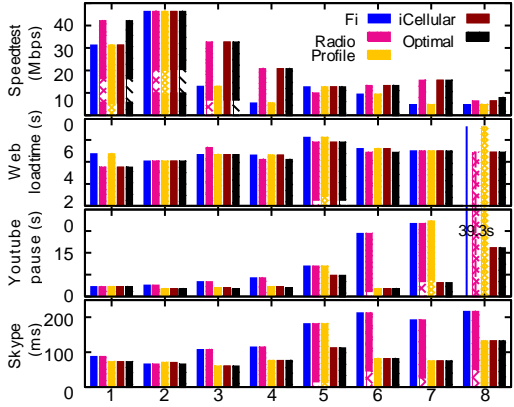

**Figure 12: Performance of speedtest, web, youtube, skype using various multi-carrier access schemes.**

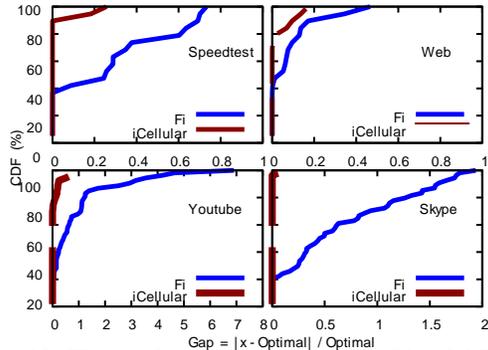

**Figure 13: The performance gaps from *Fi* and *iCellular* to the optimal.**

We compare *iCellular* and its variants, with two baselines: *(i)* the standard PLMN selection (also used in Project **Fi**'s auto selection mode), and *(ii)* **Optimal strategy**. We obtain the optimal access option by exhausting their performance at each location. It serves as an ideal performance bound. We implement the default *iCellular* using regression tree based on radio+profile, as well as other two strategies. (1) **Radio-only**: this is the *de-facto* handoff strategy in 3G/4G network. We implement the standardized cell re-selection algorithm [10, 13]: whenever there exists 4G whose signal strength is higher than -120dBm, the strongest 4G carrier network is chosen. Otherwise we choose the strongest 3G carrier network. (2) **Profile-only**: this strategy explores the heterogeneity between carriers. It moves the device to the carrier network with highest QoS (see Table 2).

**Result:** Figure 12 shows their performance in eight instances (locations), which belong to three categories: both carriers with acceptable coverage (cases 1-2), one carrier with acceptable coverage but the other not (cases 3-5), both carriers with weak coverage and one is weaker (cases 6-8). We further compare their performance with the optimal one. Let $I$ and $I_{opt}$ be the access options chosen by the test scheme and the optimal, and let $x$ and $x^*$ be their corresponding performance (per app). We define the hit ratio as the matching samples $|(I \doteq I_{opt})|$ over all the test samples. We define the gap ratio $\gamma = \frac{|x-x^*|}{x}$ and let $\gamma^+$ be the positive ratio set $\gamma^+ = \{\gamma | \gamma > 0\}$. We plot CDF of the $\gamma$ in Figure 13 and present the hit ratio and statistics of $\gamma^+$ in Table 3. We make three observations.

|  | Fi | | | iCellular | | |
|---|---|---|---|---|---|---|
|  | hit-ratio (%) | med($\gamma^+$) (|x−x*|/) | max($\gamma^+$) (|x−x|/) | hit-ratio (%) | med($\gamma^+$) (|x−x|/) | max($\gamma^+$) (|x−x|/) |
| **SpeedTest** (speed) | 31.6 | 36.2% 3.8Mbps | 73.3% 19.8Mbps | **89.5** | **12.4%** 1.4Mbps | 25.7% 9.8Mbps |
| (loadtime) |  | 0.5s | 2.3s |  | 0.2s | 0.7s |
| **Youtube** (Pause) | 15.1 | 55% 1.4s | 690% 28.1s | **69.8** | **18%** 0.3s | 111% 3.2s |
| **Skype** (Latency) | 18.5 | 62.9% 64ms | 193.8% 117ms | **98.1** | **2.5%** 4.4ms | 6.7% 4.5ms |

**Table 3: Statistics of performance gaps from the optimal one in all the tests.**

First, *iCellular* makes a wiser multi-carrier access decision. It matches with the optimal one in most cases. The hit ratios are as high as 89.5%, 68.4%, 69.8% and 98.1% in the all SpeedTest, Web, Youtube and Skype tests, respectively (Table 3). They are relatively small in case of Web and Youtube but does not incur much performance degradation (explained later). They are much higher than the ones using the standard PLMN selection procedure (below 37%). It indicates that *iCellular* overcomes the identified problems (P1, P2 and P3) and does not miss the the best-quality access in most cases.

Second, *iCellular* greatly boosts data service performance. Compared with *Fi*, *iCellular* not only narrows its performance gap (*e.g.*, reducing the maximal speed loss from 73.7% (19.7Mbps) to 25.7%, and the maximal video suspension time gap from 28.1s to 3.2s), but also achieves similar performance as the optimal one. It either hits the optimal decision or exhibit small gaps from the optimal one, regardless of various applications and coverage types. The performance gain varies with locations: with acceptable coverage (case 1-2), *Fi*'s performance also approximates the optimal one. But at locations with weak coverage, *iCellular* improves the device performance more significantly. We also notice that the performance gain varies with applications (traffic patterns). Compared with other traffic, *iCellular* provides relatively small improvement for web browsing. The reason is that, web traffic volume is relatively small and there is no large performance distinction using various access options. However, for heavy traffic like file transfer (speedtest), video streaming and voice calls, *iCellular* can significantly improves the performance. Roughly, the average improvement of *iCellular* over *F*i approximates $\gamma_{fi} - \gamma_{icellular}$. On average, *iCellular* increases 23.8% downlink speed and reduces 7.3% loading time in Web, 37% suspension time in Youtube, 60.4% latency in Skype. Since *iCellular* often selects the optimal access, the maximal gain over *Fi* can be up to 46.5% in Web, 6.9x in Youtube, 1.9x in Skype, and 3.74x in speed.

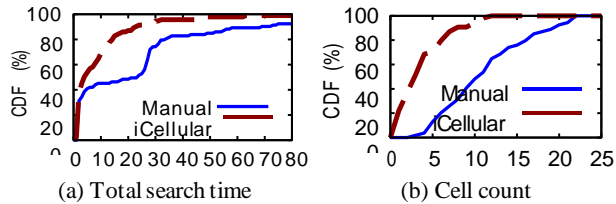

(a) Total search time  (b) Cell count

**Figure 14:** *iCellular*'s adaptive monitoring avoids exhaustive search.

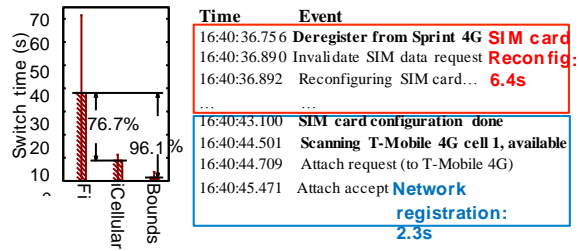

**Figure 15:** Switch time.  **Figure 16:** Implementation overhead in *iCellular*'s direct switch.

Third, *iCellular*'s regression tree algorithm best approximates the optimal strategy. It outperforms radio-only and profile-only variants (§4.4). We also notice that the importance of profile and radio measurements vary across applications. For example, our log event analysis shows that, T-Mobile assigns *Project-Fi* users with interactive traffic class (Table 2), which is optimized for delay-sensitive VoIP service [8][2]. Instead, Sprint only allocates best-effort background traffic class to users. This explains why for Skype, the profile-only strategy's performance approximates to the optimal strategy.

## 6.2 Efficiency and Low Overhead

We next present the micro-benchmark evaluation of key components and validate that *iCellular* is efficient. This partly contributes to the nice performance of *iCellular* in §6.1. We further examine its fault prevention and low overhead regarding CPU, memory and battery usage.

**Efficiency.** We examine *iCellular*'s efficiency through two adaptive module tests. First, we show *iCellular*'s adaptive monitoring is able to accelerates the carrier scanning. We compare it with the default manual search and show the total search time and the number of cells scanned at 100 different locations. Figure 14 shows that, with adaptive search, 70% of the search can be finished within 10s, which is 64% shorter than the exhaustive manual search. Figure 14b validates that such saving comes from the avoidance of scanning unnecessary cells. The search time and the number of cells vary with locations, depending on the cell density.

Second, we examine how *iCellular*'s adaptive switch can reduce service disruption. In this experiment, we place the phone at the boundary of two carrier's coverage, and test the switch time needed for *iCellular* and the PLMN selection (aka, *Project-Fi*) for 50 runs. The inter-carrier switch time is defined as the duration from the deregistration from the old network carrier, to the registration to the new carrier. For comparison, we also calculate the lower bound based on the MobileInsight event logs, described in §4.3. Figure 15 shows that *iCellu-*

---
[2]This QoS is specific to *Project-Fi*. For example, we verify that T-Mobile users with Samsung S5 is allocated with background class.

*lar* saves 76.7% switch time on average, compared with *Project-Fi*. However, the current *iCellular* implementation has not achieved the minimal switch time; it still requires 8.8s on average. We dig into the event logs to analyze the root cause. Figure 16 discloses that the current bottleneck lies in the SIM card reconfiguration. The current *iCellular* implementation relies on *Project-Fi*'s system service (universal SIM card) and it has to wait until the SIM card is reconfigured to switch to another carrier. In the experiments, we find that most of switch time (7.3s on average) are spent on the SIM card reconfiguration, which is beyond the control of *iCellular*. The phone has no network service in this period: it deregisters from the old carrier, but does not register on the new carrier. The lower bound implies that, *iCellular* could save 96.1% switch time compared with the PLMN selection, if with faster SIM card reconfiguration.

**Fault prevention.** We next verify that *iCellular* handles fault scenarios and prevents users from switching to unwise carrier networks. All three types of fault decision scenarios (§4.5) have been observed in our one-month user study of *Project-Fi* (no *iCellular* enabled). Note that the scenarios are not common (the cellular network system is largely successful). We observe one instance of the forbidden access, where one Sprint 4G base station sets the access barring option for 10 min (possibly under maintenance). *iCellular* leverages the active monitoring function and detect it from the RRC SIB1 message. Afterwards, it excludes Sprint 4G from the candidate pool, and prevent the user from making a fault decision. We observe another instance of Figure 9, where T-mobile 4G is available but T-Mobile 3G is not available. Since T-Mobile 4G does not implement Voice over LTE (VoLTE) and has to count on its 3G network (using circuit-switching Fallback) for voice calls [45]. So a correct decision should not switch to T-Mobile 4G since voice calls are not reachable. *iCellular* can detect it from the profiled voice preference and location update messages, and exclude this access option from the candidate list. We also observe uncoordinated mobility rules between network and user's strategy (see Figure 10). We validate that *iCellular* can detect conflicting scenarios and avoid such mistakes.

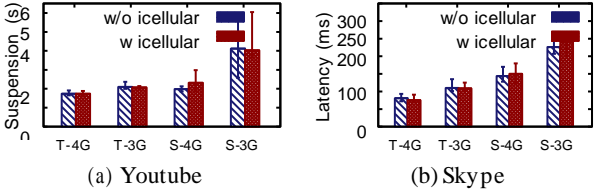

Figure 17: *iCellular*'s active monitoring has minor impacts on data performance.

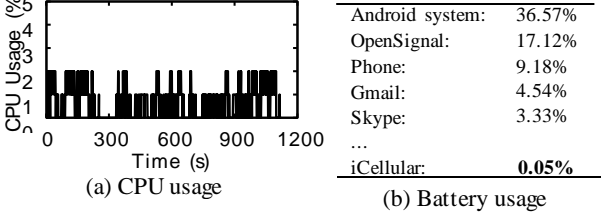

Figure 18: CPU and battery usage of *iCellular*.

**Low overhead.** We show that *iCellular* in background mode does not interrupt user's ongoing data services. Most of time *iCellular* only performs active monitoring and the switch occurs only when a new decision is made. We repetitively call Monitor() to scan the carrier networks to test its impact on the ongoing apps. We test with four apps and the results with/without *iCellular*'s monitoring are similar. Here, we only show the results for Youtube and Skype in Figure 17.

We then look into its CPU and memory usage. In all the tests, the maximum CPU utilization is below 2%, while its maximum memory usage is below 20 MB (including virtual memory). Figure 18a shows a 20-min log in a driving test (with a higher CPU/memory usage than the static and walking counterparts), where its maximum memory usage is 16.45MB. We finally measure *iCellular*'s energy consumption. We use a fully-charged Nexus 6 and run it for 24 hours, and use Battery doctor [1] to record power consumption for each component/app. Figure 18b shows one record, where *iCellular* explicitly consumes 0.05% of battery. We are aware that the current approach may not be accurate. We are still working on more tests to quantify its actual energy cost. Given the current result, we guess that the additional energy cost is not heavy. Its energy consumption can be further optimized (*e.g.* sleep mode).

## 7 What About a Clean-Slate Design?

In this section, we discuss how a clean-slate design (*e.g.* 5G) could help overcome the limitations in *iCellular*.

**Simultaneous access to multiple carriers:** although *iCellular* does not mandate simultaneous cellular access for incremental deployability (§4.1), in long term it is still preferred to give users this flexibility. For the network, the account (HSS [9]) should maintain multiple carrier contexts, and prevent users from automatic de-registration from old carriers. For the mobile device, its cellular interface should be enhanced to store multiple carrier network states. Moreover, it should have round-robin mechanisms similar to proposals in WiFi [22, 33].

**Enhanced cellular access abstractions:** with simultaneous cellular access, *iCellular*'s cellular access strategy APIs (§4.1) can be extended in two dimensions. First, the SwitchTo() function can be generalized to addition/deletion of registered carriers. Second, with simultaneous access to multiple carriers, the Predictor() function can be enhanced with runtime probing. For registered carriers, users can monitor all of their cellular events. For unregistered carriers, the Predictor() function is still identical to *iCellular*'s.

## 8 Related Work

In recent years, exploring multiple cellular carriers attracts research efforts on both network and device side. The network side efforts include sharing the radio resource [25, 32, 41, 41] and infrastructure [19, 20, 34, 48] between carriers, which helps reduce deployment cost. On the device side, both clean-slate design with dual SIM cards [2, 23] and single universal SIM card [16, 27, 29] are explored for multi-carrier access. Our work complements the single-SIM approach for incremental deployment, but differs from recent efforts by moving beyond the network-controlled roaming, and offering user-defined selection in a responsive and non-disruptive way.

*iCellular* explores the rich cellular connectivities on mobile device. Similar efforts explore the multiple physical interfaces from WiFi and cellular network, including WiFi offloading [17, 24, 26] and multipath-TCP [40, 47]. *iCellular* differs from them since it uses single cellular interface. In the WiFi context, recent works [18, 22, 33] propose aggregate multiple APs for higher capacity. As discussed in §4.1, similar techniques are unavailable for 3G/4G. Instead, *iCellular* chooses to let users customize the selection strategies between carriers.

## 9 Conclusion

The current design of cellular networks limits the user's ability to fully explore multi-carrier access. The fundamental problem is that, existing 3G/4G (even 5G) mobile networks place most decisions and operational complexity on the infrastructure side. This network-centric design is partly inherited from the legacy telecom-based paradigm. As a result, the increasing capability of user devices is not properly exploited. In the multi-carrier access context, users may suffer from low-quality access while incurring unnecessary service disruption. In this work, we describe *iCellular*, which seeks to leverage the end device's intelligence to dynamically select better carrier through adaptive monitoring and online learning. Our initial evaluation seems to partially validate the viability of our approach.

# Appendices

## A  Cellular Access Strategy for Minimizing Data Billing

If the users have contracts with multiple carriers with different data charging plans(*e.g.* Apple SIM), they can minimize their data charging with the following access strategy. The user first profile the data charing plan from each carrier network, which tend to be static(*e.g.* [5, 6]). Note that some carrier's charging depends on user's current data usage. So the user monitors the uplink/downlink data usage, and pre-computes the per-unit data charging for each carrier network. Then the user chooses the carrier network with the minimal per-unit data charging, and asks *iCellular* to switch to that. Simple calculation shows that this minimizes user's total billing.

## B  Collecting App-specific Performance

For SpeedTest, we directly record the downlink speed for each test. Note that Nexus 6 supports LTE category 4, which can yield up to 150Mbps downlink bandwidth in theory [7]. This is why we observe 40+Mbps downlink speed in our tests, which is much higher than most previous measurements.  For Web, we run Firefox to to fetch yahoo news webpage (http://news.yahoo.com) 5 times, and get the page loading time from Firefox's debugging console [37]. For Youtube, we extract its buffering time by tracking OnBuffer(True) and OnBuffer(False) events from Youtube Android player API [28], and calculating the elapsed time in between, during which the user has to pause the video. For Skype, we collect round-trip latencies (in ms) as the performance metric. To get it, We enabled the Technical info panel in the Skype app, which shows the latency in the call. For each test, we run a 2-min VoIP call and record the round-trip latency in every second. Then we calculate the median latency in the call.